\documentclass{appolb}
\usepackage{epsfig}
\usepackage{cite}

\begin{document}
\pagestyle{plain}
\title{\bf HIGGS ${\cal CP}$ FROM {\boldmath  $H/A^0\to\tau \tau$ DECAY}
\thanks{Presented at the Cracow Epiphany Conference on {\it Heavy Flavors}, 
3-6 January 2003, Cracow. } 
~\thanks{{\bf Report-no:} TTP03-13 }
}
\author{Ma\l gorzata Worek
\address{
Institut f\"{u}r Theoretische Teilchenphysik (TTP) \\
Universit\"{a}t Karlsruhe,
D-76128 Karlsruhe, Germany.
\\Institute of  Physics, University of Silesia \\ Uniwersytecka 4,
 40-007 Katowice, Poland. \\ e-mail: {\tt Malgorzata.Worek@ifj.edu.pl}}
}
\maketitle

\begin{abstract}

We show how the transverse $\tau^{+}\tau^{-}$ spin correlations can be
used  to measure the parity of the Higgs boson  and hence to
distinguish a ${\cal CP}$-even $H$ boson from ${\cal CP}$-odd $A^{0}$
in the future high energy accelerator experiments. We  investigate
the subsequent decays of the $\tau^{\pm}$ into
$\pi^{\pm}\bar{\nu}_{\tau}(\nu_{\tau})$ and
$\rho^{\pm}\bar{\nu}_{\tau}(\nu_{\tau})$.  The prospects for the
measurement of the Higgs boson parity with a mass  of $120$ $GeV$ are
quantified for the case of $e^{+}e^{-}$ collisions of  $500$ $GeV$
center of mass energy.  The Standard Model Higgsstrahlung production
process is used as an example.
\end{abstract}
\section{Introduction}

In the Standard Model (${\cal SM}$) of elementary particle
interactions, the  breaking of electroweak symmetry is achieved through the
Higgs mechanism.  The simplest realization is provided by the
introduction of a complex  Higgs doublet, which leads to the presence
of a neutral ${\cal CP}$-even  Higgs boson in the physical
spectrum. Among the possible extensions of the ${\cal SM}$, the
Minimal Supersymmetric Standard Model (${\cal MSSM}$) has  been
considered most seriously. The minimal realization of the Higgs
mechanism within supersymmetric extensions of the standard model
requires the presence of two Higgs doublets at low energies. After the
Higgs mechanism operates, five real fields remain, and there should be
five spin zero Higgs fields, and the spectrum includes also a pseudoscalar
${\cal CP}$-odd state. This non trivial assignment of  the quantum
numbers requires the investigation of experimental  opportunities to
measure the parity of the Higgs boson states.
The investigation of the mechanism of the electroweak symmetry
breaking is one of the central task of a future  $e^+e^-$ linear
collider operating at center-of-mass energies between  $350$ and
$1000$ $GeV$. This accelerator will allow to study  completely
and with high accuracy also the profile of the Higgs sector.

The problem of
the Higgs boson(s) parity measurement  was approached  in general form
quite early in Ref.\cite{Dell'Aquila:1988rx,Dell'Aquila:fe}  for Higgs
decays into fermions and gauge boson pairs. It has recently been
revived \cite{Barger1994,Hagiwara1994,Skjold:1994qn,Boe:1998kp,Hagiwara2000,Kramer:1994jn,Grzadkowski:1995rx}.

In this study we investigate the case of a Higgs boson which is light
enough that the $W^+W^-$ decay channel remains closed.  Then the most
promising decay channel of neutral Higgs particles is the
$b\bar{b}$ channel, ${\cal BR}(H\to b\bar{b})\sim$ $90\%$. 
This applies in the  ${\cal SM}$ as well as in the
minimal supersymmetric extension like  ${\cal MSSM}$
\cite{Aguilar-Saavedra:2001rg}. However, due to depolarization
effects in the fragmentation process, it is very difficult to extract
information on the $b$ polarization state \cite{Mele:1992kh}. A much
cleaner  channel, though with branching ratios suppressed by an order
of magnitude,  
is the $\tau^+\tau^-$ mode,
 ${\cal BR}(H\to \tau^{+}\tau^{-})\sim$ $9\%$.  The $\tau^+\tau^-$
channel  is useful in the ${\cal SM}$ for Higgs masses less than $\sim 140$
$GeV$. Up to this mass, the Higgs particle is very narrow,
$\Gamma(H)\le 10$ $MeV$.  In Supersymmetric theories, the  $\tau^+\tau^-$
channel is useful over a much larger mass range.
The main production mechanism of the ${\cal SM}$ Higgs
boson in $e^{+}e^{-}$ collisions in the future Linear Collider ($LC$)
are the Higgsstrahlung process, $e^{+}e^{-}\to ZH$ and the $WW$ fusion 
process,  $e^{+}e^{-}\to W^{*}W^{*} \to \bar{\nu}_{e}\nu_{e}H$. 
The cross section for the 
Higgsstrahlung process scales as $\sim 1/s$ and dominates at low energies 
while the cross section for the $WW$ fusion 
process rises as $\sim \log (s/m^{2}_{H})$ and dominates at high 
energies \cite{Aguilar-Saavedra:2001rg}. At $\sqrt{s}=500$ $GeV$,
the Higgsstrahlung and the 
$WW$ fusion processes have approximately the 
same cross section for $100$ $GeV$ $\le m_{H}\le 200$ $GeV$.

In our analysis, we will 
take as an example the $e^{+}e^{-}\to ZH$; 
$Z\to \mu^{+} \mu^{-}$; $H\to \tau^{+} \tau^{-}$   
production process. We discuss a method for the parity
measurement of
the Higgs boson  with a mass  of $120$ $GeV$ for the case 
of $e^{+}e^{-}$ collisions with  $\sqrt{s}=500$ $GeV$
using  $H/A^0\to\tau^+\tau^-$;
$\tau^\pm\to\pi^\pm\bar{\nu}_{\tau}(\nu_{\tau})$ and
$\tau^\pm\to\rho^\pm\bar{\nu}_{\tau}(\nu_{\tau})$;
$\rho^\pm\to\pi^\pm\pi^0$ decay chains.  
All the Monte Carlo samples have been
generated  with the {\tt TAUOLA}  
library \cite{Jadach:1990mz,Jezabek:1991qp,Jadach:1993hs}.  
For the production of the $\tau$ lepton pairs the Monte Carlo program   
{\tt PYTHIA 6.1}  is used \cite{Pythia}. 
The effects of initial state   
bremsstrahlung were included in the {\tt PYTHIA} generation.  
For  the $\tau$ lepton pair decay   
with full spin effects included in the $H\to\tau^{+}\tau^{-}$; 
$\tau^{\pm}\to\pi^{\pm}\bar{\nu}_{\tau}(\nu_{\tau})$ and 
$\tau^{\pm}\to\rho^{\pm}\bar{\nu}_{\tau}(\nu_{\tau})$; 
$\rho^{\pm}\to\pi^{\pm}\pi^{0}$ chains, the  interface explained in  
Refs.\cite{Was:2002gv,Bower:2002zx} was used. It is an extended version of 
the standard universal interface 
of Refs.\cite{Pierzchala:2001gc,Worek:2001hn}. 

The rest of the paper is
organized as follows. In section 2 the theoretical considerations 
used to understand the decay chain of the Higgs particle are explained.
In section 3 and 4 we define the observables we use to distinguish between
the scalar and pseudoscalar Higgs boson in the 
$\tau^\pm\to\pi^\pm\bar{\nu}_{\tau}(\nu_{\tau})$ and 
$\tau^\pm\to\rho^\pm\bar{\nu}_{\tau}(\nu_{\tau})$ decays respectively. 
In these sections we list our assumptions on detector effects 
and the imposed cuts as well as the main numerical results. 
A summary closes the  paper. 
\section{Higgs boson parity}
The $H/A$ parity information must be extracted from the correlations
between $\tau^{+}$ and $\tau^{-}$ spin components which are further
reflected in correlations between the $\tau$ decay products in the
plane transverse to the $\tau^{+}\tau^{-}$ axes.  This is because  the
decay probability, see {\it e.g.}  Ref.~\cite{Kramer:1994jn},
\begin{equation}
\Gamma(H/A^0\to \tau^{+}\tau^{-}) \sim 1-s^{\tau^{+}}_{\parallel}
s^{\tau^{-}}_{\parallel}\pm s^{\tau^{+}}_{\perp}s^{\tau^{-}}_{\perp}
\label{densi}
\end{equation}
is sensitive to the $\tau^\pm$ polarization vectors $s^{\tau^{-}}$ and
$s^{\tau^{+}}$  (defined in their respective rest frames, the $z$-axis
is oriented  in the $\tau^{-}$ flight direction).  The symbols
${\parallel}$/${\perp}$ denote components parallel/transverse to the
Higgs boson momentum as seen from the respective $\tau^\pm$  rest
frames.  This suggests that the experimentally clean
$\tau^{+}\tau^{-}$ final state may be the proper instrument to study
the parity of the Higgs boson.  

Now, a few representative examples of the $\tau$ lepton decay will be
discussed  in more detail. In particular, we analyze the $\tau$ decay
into one and  two pions.

\section{Higgs $CP$ from $\tau^\pm\to\pi^\pm\bar{\nu}_{\tau}(\nu_{\tau})$ 
decay}
Exploring transverse spin correlations in the Higgs
boson decay   $H/A^{0} \to \tau^+ \tau^-$;  $\tau^\pm\to \pi^\pm
\bar{\nu}_{\tau}(\nu_{\tau})$ one can provide a model
independent parity  determination  test. The method relies  
on the properties of the Higgs
boson Yukawa coupling to the $\tau$ lepton,  which in the general case
can\ be written as  $ \bar\tau (a_\tau + i b_\tau \gamma_5) \tau$. The
method does not depend on the Higgs boson production mechanism at all.
Even though the $\tau^{\pm}\to\pi^{\pm}\bar{\nu}_{\tau}(\nu_{\tau})$ 
decay mode is rare,
${\cal BR}(\tau \to \pi \nu_{\tau})\sim 11\%$, it may serve as a  simple
example that illustrates the basic principles.

The spin of the $\tau^{\pm}$ leptons
is not directly observable but  translates directly into correlations
among their decay products. To demonstrate this we 
define the polar angles between the $\pi^{\pm}$ and the $\tau^{-}$
direction in the $\tau^{\pm}$  rest frames by $\theta^{\pm}$ and the
relative azimuthal angle $\phi^{*}$ between the decay planes.
The angular correlation between $\tau^{+}\tau^{-}$ decay products 
may then be written as \cite{Kuhn:1982di}
\begin{equation}
\frac{1}{\Gamma}\frac{d\Gamma(H/A^{0}\to
\pi^{+}\bar{\nu}_{\tau}\pi^{-}
\nu_{\tau})}{d\cos\theta^{+}d\cos\theta^{-}d\phi^{*}}=\frac{1}{8\pi}
\left[ 1+\cos\theta^{-}\cos\theta^{+}\mp \sin\theta^{+}\sin\theta^{-}
\cos\phi^{*} \right].
\label{equation1}
\end{equation}

A simple asymmetry in the azimuthal angle that projects out the parity
of the  particle can be derived \cite{Dell'Aquila:1988rx,Barger1994}
by integrating  out the polar angles
\begin{equation}
\frac{1}{\Gamma}\frac{d\Gamma(H/A^{0})}{d\phi^{*}}=\frac{1}{2\pi}\left[
1\mp \frac{\pi^{2}}{16}\cos\phi^{*} \right].
\label{equation2}
\end{equation}
A useful observable sensitive to the parity of the decaying Higgs
particle is the angle $\delta$ between the two charged pions in the
Higgs boson rest frame \cite{Kuhn:1982di}.
This delicate measurement  will
require, however, the reconstruction of the Higgs boson rest-frame.  
It is generally accepted  that the Monte Carlo method is the only way
to estimate whether the  measurement can be realized in practice, and
which features   of the future detection set up may turn out to be
crucial.  To enable such studies  we have extended the standard
universal interface\cite{Pierzchala:2001gc,Worek:2001hn}, 
of the {\tt TAUOLA} $\tau$-lepton decay  library
\cite{Jadach:1990mz,Jezabek:1991qp,Jadach:1993hs}, to include  the
complete spin effects for $\tau$ leptons originating  from the  spin
zero particle\cite{Was:2002gv,Bower:2002zx}.  The  
interface is expected to work with any Monte
Carlo generator providing Higgs boson production, and subsequent decay
into a pair of $\tau$ leptons.
\begin{figure}[!ht]
\begin{center}  
\epsfig{file=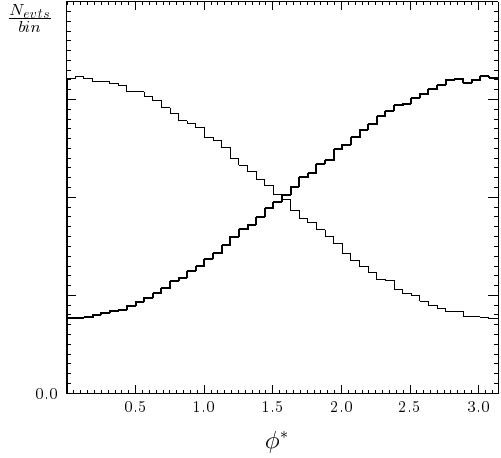,width=80mm,height=70mm}
\end{center} 
\caption  
{\it The $\pi^+ \pi^-$ acoplanarity  distribution (angle $\phi^*$)  
in the  Higgs boson rest frame.   The thick line denotes the case 
of the scalar Higgs boson and 
thin line the pseudoscalar one.}
\label{rysunek1}  
\end{figure}  
\begin{figure}[!ht]
\begin{center}  
\epsfig{file=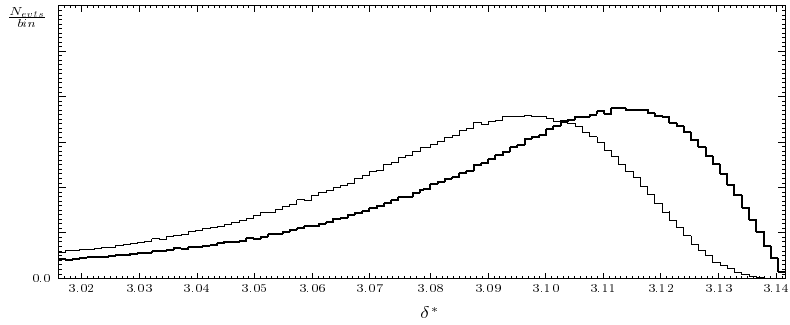,width=130mm,height=60mm}
\end{center} 
\caption  
{\it The  $\pi^+ \pi^-$ acollinearity distribution  (angle $\delta^*$)  
in the  Higgs boson rest frame.
Parts of the distribution close to the end of the spectrum; 
$\delta^* \sim \pi$ are shown. No  smearing is done. 
The thick line denotes the case 
of the scalar Higgs boson and the
thin line the pseudoscalar one.
}
\label{rysunek3}  
\end{figure}  

Let us
turn to the discussion of numerical results. As an example we took a
Higgs boson of $120$ $GeV$.
Fig.~\ref{rysunek1}  presents the distribution in the angle $
\phi^*=\arccos ({\vec n}^{+} \cdot {\vec n}^{-})$ where
\begin{equation}  
{\vec n}^{\pm}={ {\vec p}^{~\pi^\pm} \times  \; {\vec p}^{~\tau^-}
\over |  {\vec p}^{~\pi^\pm} \times \;  {\vec p}^{~\tau^-}|},
\end{equation}
{\it i.e.} the  acoplanarity angle. Thick lines will denote
predictions for the scalar Higgs boson and  thin lines for the
pseudoscalar one. The  distribution
is indeed, as it should be,  proportional to  $\sim 1 \mp {\pi^2 \over
16} \cos \phi^*$ respectively for scalar and pseudoscalar Higgs boson, see
Eg.(\ref{equation2}).  In Fig.~\ref{rysunek3} we plot the
distribution of the $\pi^+ \pi^-$  acollinearity angle  ($\delta^*$).
The difference between the case of a scalar and a pseudoscalar Higgs
boson is  clearly visible, especially for acollinearities close to $\pi$.

To test the feasibility of the measurement, some assumptions about the
detector effects have to be made.   We include, as the most critical
for our discussion, effects due to    inaccuracies in the measurements
of the   $\pi^{\pm}$ momenta.  We assume Gaussian spreads of the
`measured' quantities with    respect to the generated ones. For
charged pion  momentum we assume a 0.1\% spread on its energy
and direction.

If the information on the beam energies and energies of all
other observed particles (high $p_T$ initial state bremsstrahlung
photons, decay products of $Z$ {\it etc.}) are taken into
consideration the Higgs boson rest frame  can be reconstructed.  We may
define the ``reconstructed'' Higgs boson  momentum as the difference
of the sum of the beam energies and the  momenta of all visible 
particles, that is decay  products of the $Z$ and all radiative 
photons of $|\cos\theta|< 0.98$. In our study we will mimic 
in a very crude way the beamstrahlung
effects, assuming only a  flat  spread over the range  of $\pm$
$5$ $GeV$ for the longitudinal component of the Higgs boson momentum
with respect to the generated one\footnote{The typical spread  for the beam 
energy in a linear collider is of the order of a 
few percent\cite{Abe:2001gc}.}.

\begin{figure}[!ht]
\begin{center}  
\epsfig{file=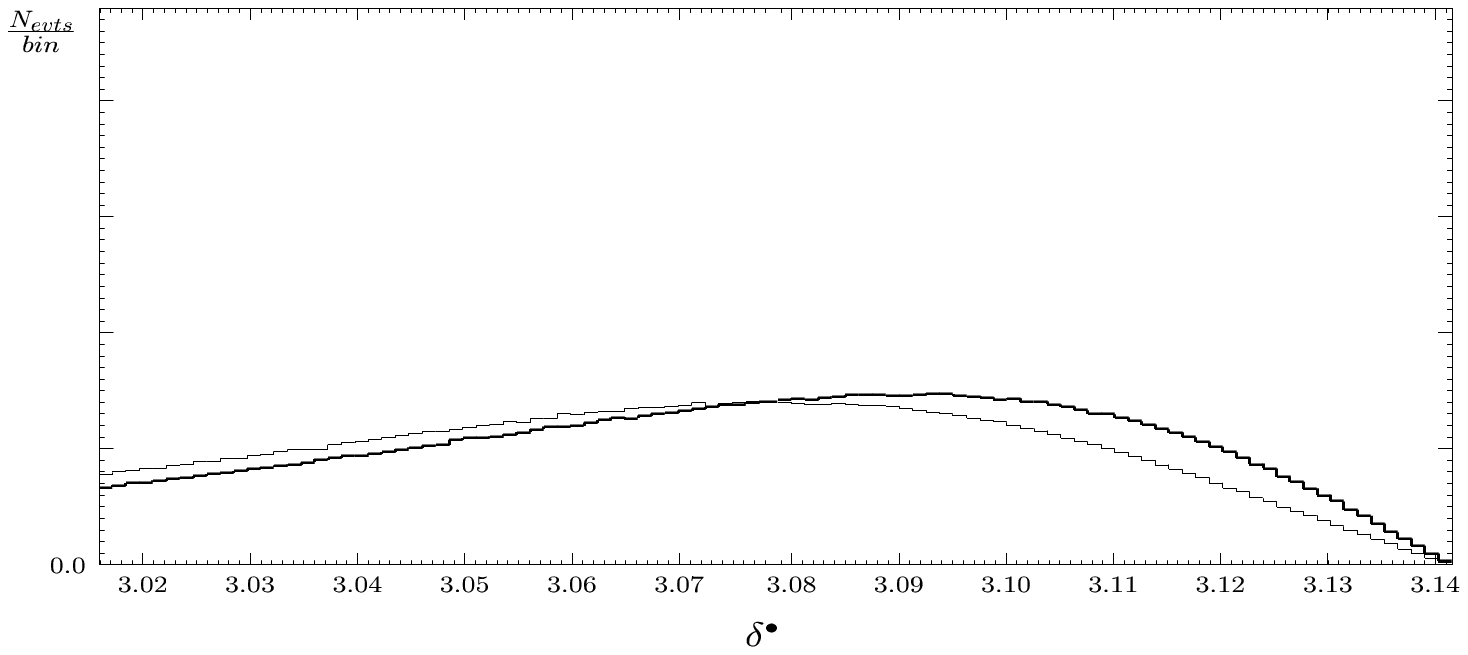,width=130mm,height=60mm}
\end{center} 
\caption  
{\it The  $\pi^+ \pi^-$ acollinearity distribution  (angle $\delta^\bullet$)  
in the  reconstructed Higgs boson rest frame.  All smearing is included.
Parts of the distribution close to the end of the spectrum; 
$\delta^\bullet \sim \pi$ are shown.
The thick line denotes the case 
of the scalar Higgs boson and the
thin line the pseudoscalar one.
}
\label{rysunek6}  
\end{figure}  

Fig.~\ref{rysunek6} shows us
the distribution of the acollinearity  angle ($\delta^\bullet$) build
from smeared $\pi^{\pm}$ momenta defined in the reconstructed Higgs boson 
rest-frame. The difference between the scalar and the pseudoscalar Higgs 
bosons is only  barely visible. We
have studied several mechanisms of Higgs boson production, in all
cases depletion  of  the acollinearity distribution sensitivity to
transverse spin effect was quite similar.  We can conclude that our
results are thus independent from the production mechanism.  

The beamstrahlung effect taken into account in the 
reconstruction of the  Higgs boson four-momentum degraded the method
of measuring the Higgs boson parity using the decay chain $H \to
\tau^+\tau^-$, $\tau^\pm \to \pi^\pm \bar{\nu}_{\tau}(\nu_{\tau})$ in
a decisive way. Therefore, there is little hope for  the  elegant method to
check the Higgs boson parity using its decay to $\tau^\pm\to \pi^\pm
\bar{\nu}_{\tau}(\nu_{\tau})$,  whatever the luminosity of the future
Linear Collider is assumed, unless other, unfortunately less
sensitive, to spin, than $\tau^\pm \to \pi^\pm \bar{\nu}_{\tau}(\nu_{\tau})$, 
decay modes are used as well.

\section{Higgs $CP$ from 
$\tau^\pm\to\rho^\pm\bar{\nu}_{\tau}(\nu_{\tau})$ decay}

The discussion of the $\tau$ decay to a two pion final state follows 
very much the same line. We may treat the hadron system as a single particle 
with definite spin and mass, $m_{\rho}=\sqrt{Q^{2}}$, or we may extract 
additional information from the individual pion momenta. The first strategy 
leads to a simple generalization of the single pion case. However, the effect
is diminished as compared with the $\tau^{\pm}$ decays to single 
$\pi^{\pm}\bar{\nu}_{\tau}(\nu_{\tau})$.  
This is because for the $\tau^{\pm}\to\rho^{\pm}\bar{\nu}_{\tau}(\nu_{\tau})$
decay the angular correlation term is reduced by the factor 
$(m^{2}_{\tau}-2Q^{2})^{2}/(m^{2}_{\tau}+2Q^{2})^{2}$

\[
\frac{1}{\Gamma}\frac{d\Gamma(H/A^{0}\to
\rho^{+}\bar{\nu}_{\tau}\rho^{-}
\nu_{\tau})}{d\cos\theta^{+}d\cos\theta^{-}d\phi^{*}}=
\]
\begin{equation}
\frac{1}{8\pi}
\left[ 1+ \frac{(m^{2}_{\tau}-2Q^{2})^{2}}{(m^{2}_{\tau}+2Q^{2})^{2}}\left[ \cos\theta^{-}\cos\theta^{+}\mp \sin\theta^{+}\sin\theta^{-}
\cos\phi^{*} \right]\right] .
\label{equation2}
\end{equation}
The mass of the hadronic system, $Q^{2}$, can no longer be neglected 
relative to $m^{2}_{\tau}$\cite{Hagiwara:2002fs}.

Let us turn to the case when additional information coming from
$\rho^{\pm}$ decay products can be used directly\cite{Bower:2002zx}.
The $\tau^{\pm} \to \rho^{\pm} \bar{\nu}_{\tau}(\nu_{\tau})$  decay
is very interesting because it has, by far, the largest branching
ratio, 
${\cal BR}(\tau \to \rho \nu_{\tau})\sim 25\%$. 
The polarimetric force of this channel can be improved if information
on  the $\rho$ decay products, {\it i.e.} on details of the decay 
 $\tau^{\pm} \to \pi^{\pm} \pi^0 \bar{\nu}_{\tau}(\nu_{\tau})$, is
used. This is of no surprise  because the polarimetric vector is given
by  the formula  
\begin{equation}  
h^i =  {\cal N} \Bigl( 2(q\cdot N)  q^i -q^2  N^i \Bigr)  
\end{equation}  
where ${\cal N}$  is a normalization function,  $q$  is the difference   
of the $\pi^\pm $ and $\pi^0$ four-momenta and $N$ is the
four-momentum of  the $\tau$ neutrino (all defined in the $\tau$ rest frame)   
see, {\it e.g.} \cite{Jadach:1990mz}. Obviously, any control on the
vector  $q$ can be advantageous. It is of interest to note that  
in the $\tau$ lepton rest frame, when $m_{\pi^{\pm}}=m_{\pi^{0}}$ is   
assumed, the term   
\begin{equation}  
 q\cdot N = (E_{\pi^\pm} - E_{\pi^0}) m_\tau.  
\label{formula}  
\end{equation}  
Thus, to exploit this part of the polarimetric vector, we need to have
some  handle on the difference of the $\pi^\pm$ and $\pi^0$ energies
in their  respective $\tau^{\pm}$ leptons rest frames. Otherwise, the effect
of  this part of the  polarimetric vector cancels out and one is left
with  the part proportional to the $\rho$ (equivalently $\nu_{\tau}$)
momentum.  

\begin{figure}[!ht]
\vspace{0.9 cm}
\begin{center}  
\epsfig{file=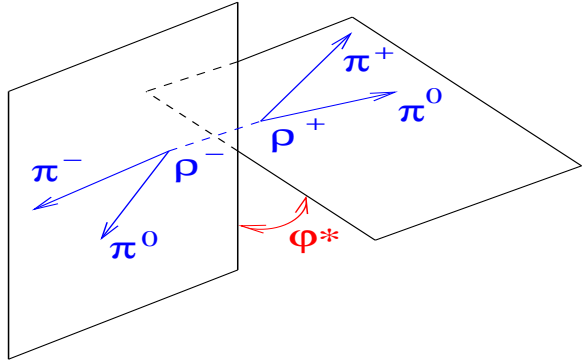,width=80mm,height=50mm}
\end{center} 
\caption  
{\it Definition of the $\rho^{+}\rho^{-}$ decay products' acoplanarity
distribution angle, $\varphi^{*}$ in the rest frame of the
$\rho^{+}\rho^{-}$ pair.  }
\label{rho-decay-products}  
\end{figure}

Let us now discuss a new   observable which we have introduced
to distinguish between the scalar and the pseudoscalar Higgs boson.
We advocate the observable where we ignore the part of the
polarimetric vector   proportional to the $\rho^{\pm}$ (equivalently
$\nu_{\tau}$) momentum in the $\tau$    rest frame.    
We rely only on the part of
the vector due to the differences of    the $\pi^\pm$ and $\pi^0$
momenta, which manifests the spin state of the $\rho^\pm$.
In the Higgs boson rest frame the $\rho$ momentum represents a larger
fraction of    the Higgs's energy than the neutrino. Therefore, we
abandon the    reconstruction of the    Higgs boson rest frame and instead
we use the $\rho^{+}\rho^{-}$ rest    frame which has the advantage
that it is built only from directly visible decay products of the
$\rho^{+}$ and $\rho^{-}$.
In the rest frame of the $\rho^{+}\rho^{-}$ system we define the
acoplanarity angle,    $\varphi^{*}$, between the two planes spanned
by    the immediate decay products (the $\pi^\pm$ and $\pi^0$) of the
two $\rho$'s, see Fig.~\ref{rho-decay-products}.

\begin{figure}[!ht]
\hspace{-0.6 cm} 
\epsfig{file=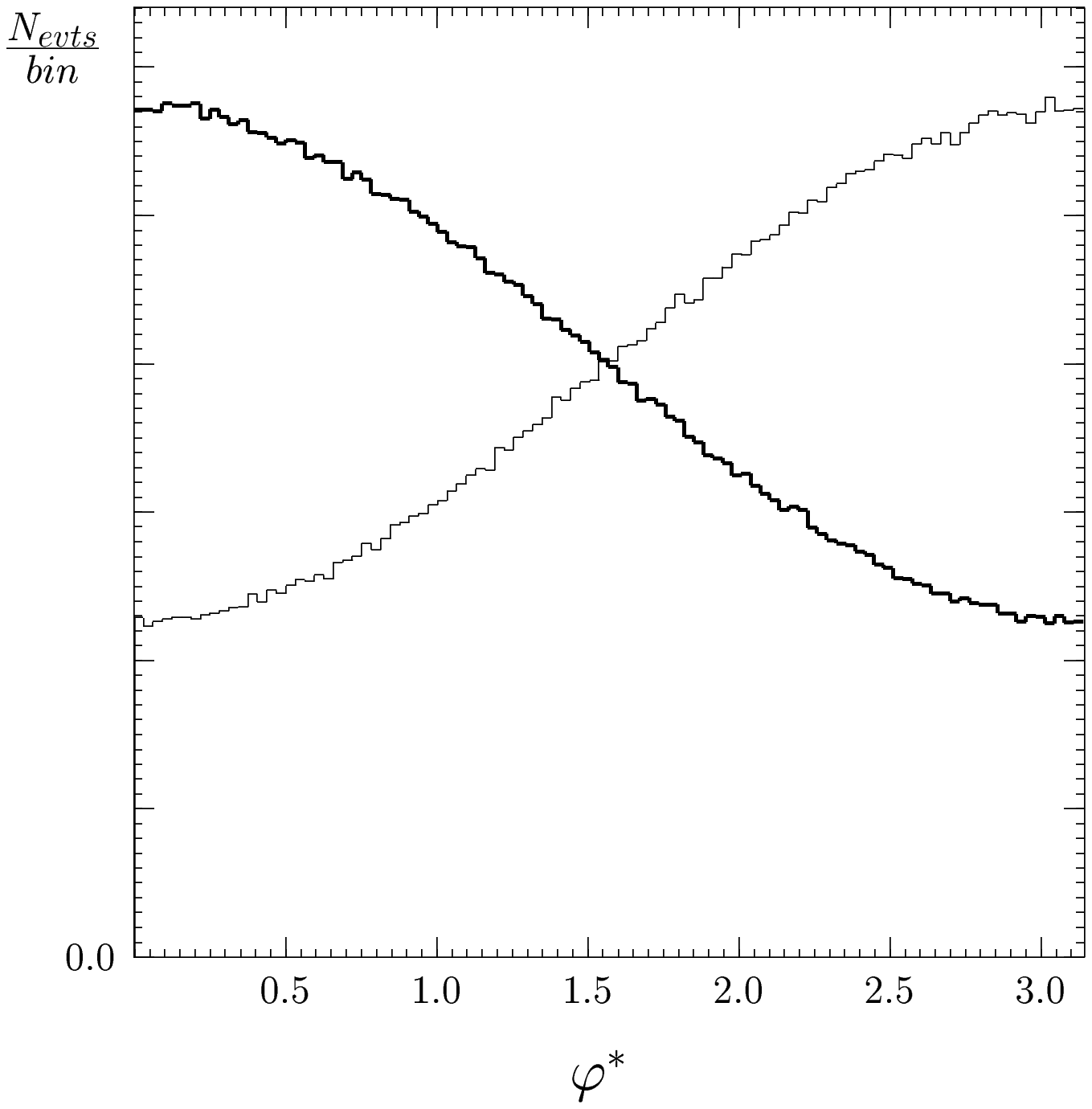,width=70mm,height=60mm}
\hspace{-0.6 cm}
\epsfig{file=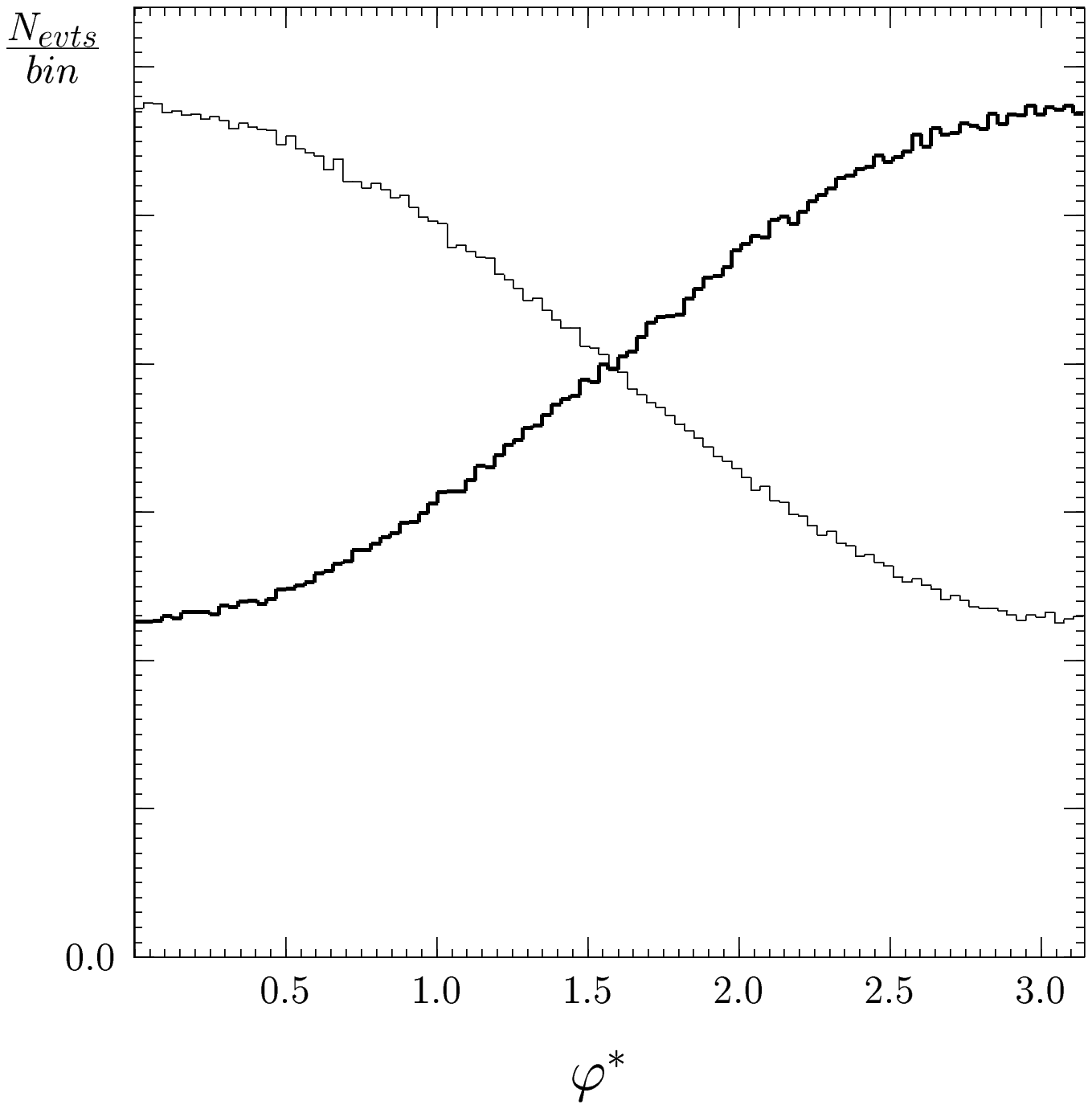,width=70mm,height=60mm}
\caption  
{\it  The $\rho^+ \rho^-$ decay products' acoplanarity distribution   
angle, $\varphi^*$, in the rest frame of the $\rho^+ \rho^-$ pair. A
cut on the differences of the $\pi^\pm$ $\pi^0$ energies defined in
their  respective  $\tau^\pm$ rest frames to be of  the same sign,
selection  $y_{1}y_{2}>0$, is used in the left plot and the
opposite sign, selection $y_{1}y_{2}<0$, is used for the right
plot. No  smearing is done.  Thick lines denote the case   
of the scalar Higgs boson and thin lines the pseudoscalar one.    }
\label{aco}  
\end{figure}  
\begin{figure}[!ht]
\hspace{-0.6 cm}
\epsfig{file=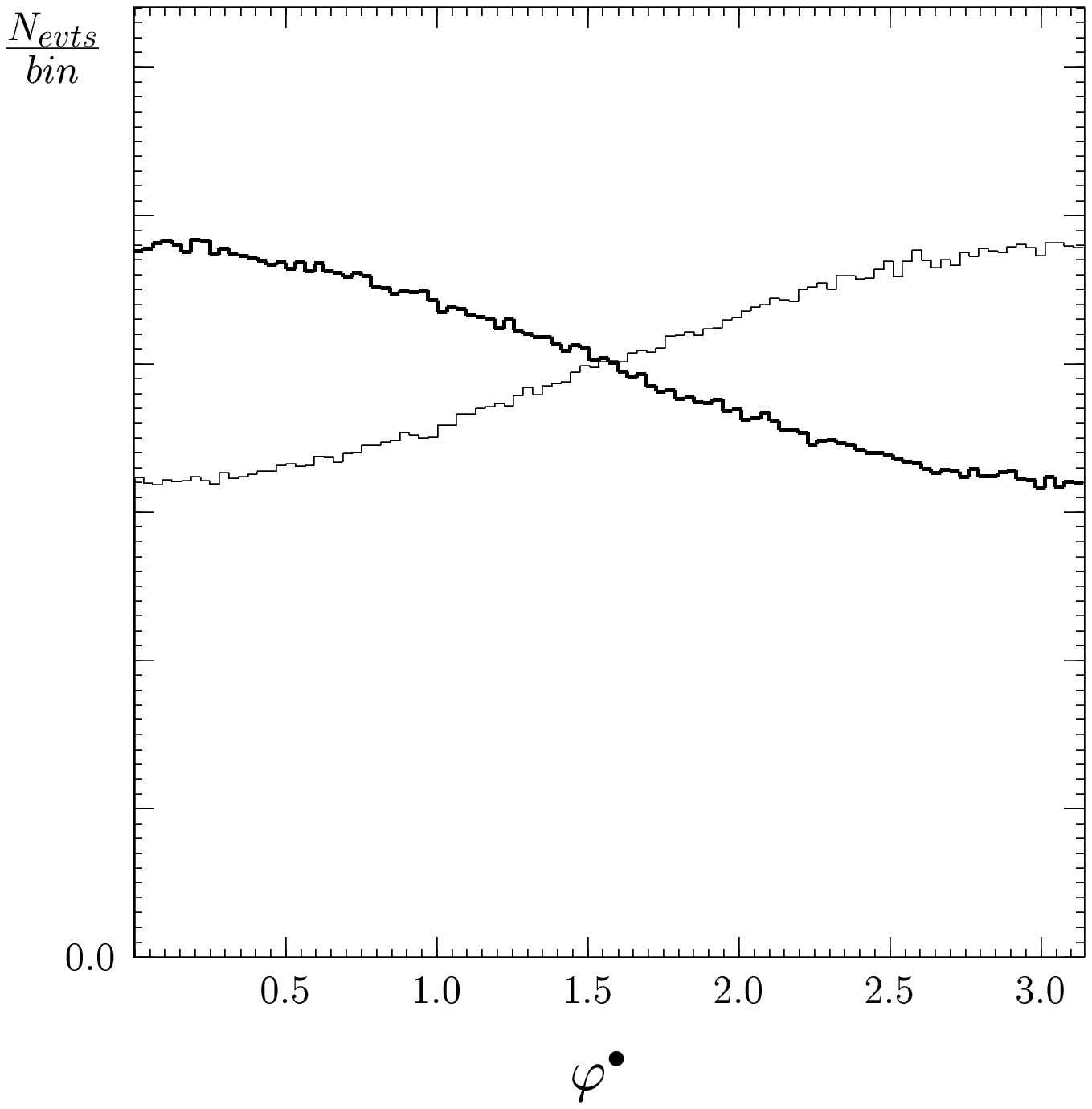,width=70mm,height=60mm}
\hspace{-0.6 cm}
\epsfig{file=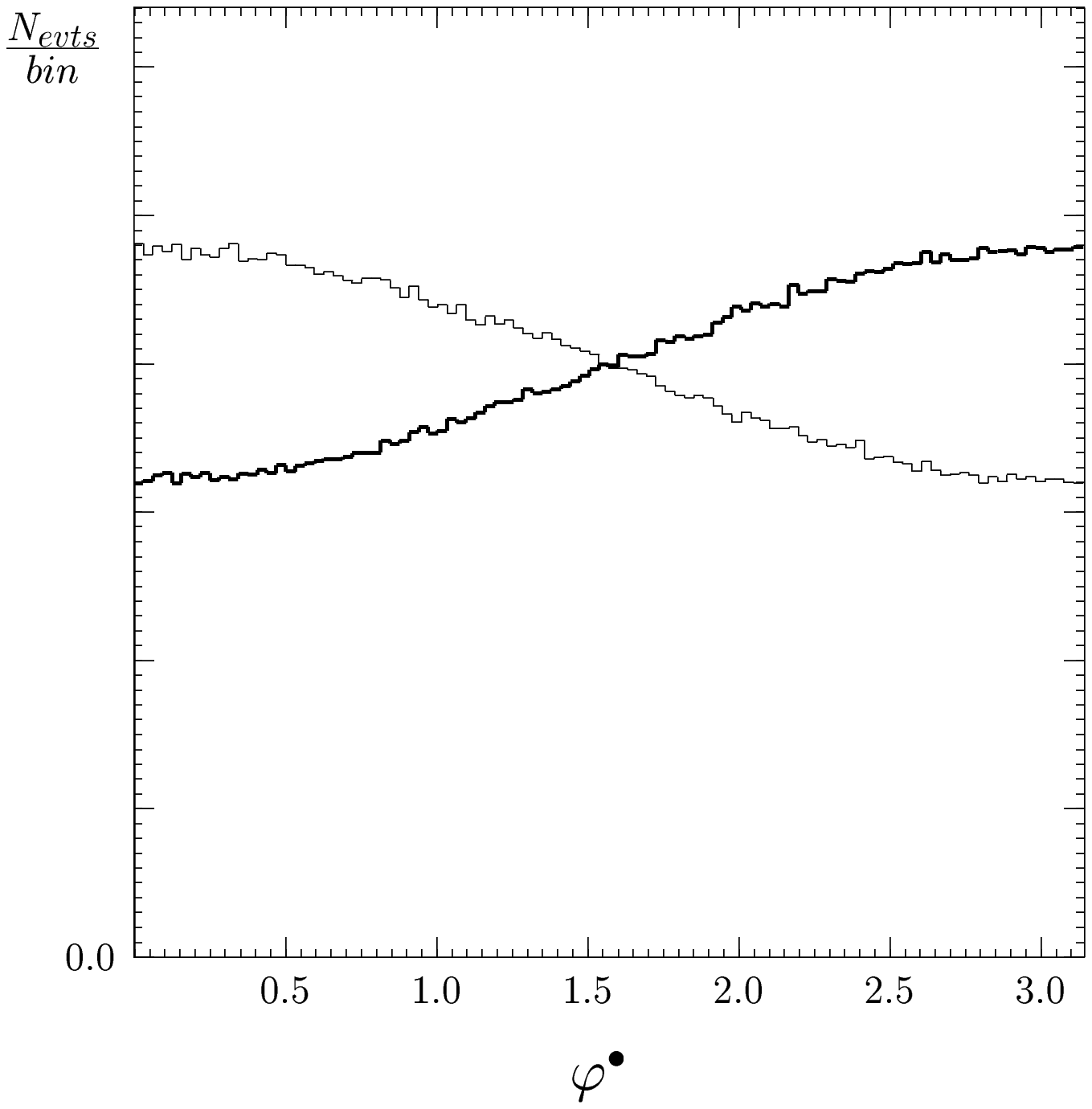,width=70mm,height=60mm}
\caption  
{\it The $\rho^+ \rho^-$ decay products' acoplanarity distribution   
angle, $\varphi^\bullet$, in the rest frame of the $\rho^+ \rho^-$
pair. A cut on the  differences of the $\pi^\pm$ $\pi^0$ energies
defined  in their respective replacement  $\tau^\pm$ rest frames to be
of   the same sign, selection $y_{1}y_{2}>0$, is used in the left
plot  and the opposite sign, selection $y_{1}y_{2}<0$, is used for
the  right plot. All smearing is included.  Thick lines denote the case   
of the scalar Higgs boson and thin lines the pseudoscalar one.    }
\label{acosm2}  
\end{figure}  

The variable $\varphi^{*}$ alone does not distinguish the scalar and
psuedoscalar Higgs boson. To do   this we must go further. The $\tau^{\pm} \to
\pi^\pm \pi^0 \bar{\nu}_{\tau}(\nu_{\tau})$ spin sensitivity    
is proportional to    the
energy difference  of  the charged and neutral pion   (in the  $\tau$
rest frame), see  formula (\ref{formula}).    We have to separate
events into     two zones, $C$ and $D$,
  
\[  
 C:~~~~~~~~~~~~~~~~  y_1 y_2>0
\]  
\[  
 D:~~~~~~~~~~~~~~~~  y_1 y_2<0
\] where,  
\begin{equation}  
y_1={E_{\pi^{+}}-E_{\pi^{0}}\over E_{\pi^{+}}+E_{\pi^{0}}}~;~~~~~
y_2={E_{\pi^{-}}-E_{\pi^{0}}\over E_{\pi^{-}}+E_{\pi^{0}}}.
\label{E-zone}   
\end{equation}\\
$E_{\pi^{\pm}}$ and  $E_{\pi^{0}}$ are    the $\pi^{\pm}, \pi^{0} $
energies in the respective    $\tau^\pm$ rest frames.  

In Fig.~\ref{aco} we plot the distribution of $\varphi^{*}$,   where
the left hand plot contains the events where the   energy difference
between the $\pi^+$ and $\pi^0$ defined in the    $\tau^+$  rest frame
is of the same sign as the energy difference of $\pi^-$ and $\pi^0$
defined in $\tau^-$  rest frame (selection  $y_1 y_2>0$).    
The right hand plot contains the events with
the opposite signs for the two   energy differences (selection
$y_1 y_2<0$). It can be seen that the    differences between
the scalar and pseudoscalar Higgs boson are large.   If the energy
difference cut was not applied, we would have completely    lost
sensitivity to the Higgs boson parity.
  
Unfortunately, since the $\tau$-lepton is not measurable,    such a
selection cut cannot be used directly. We reconstruct the $\tau$ lepton
 rest frames, and we assume Gaussian spreads of
the `measured' quantities with    respect to the generated ones.
For charged pion momentum we assume a $0.1\%$ spread on its energy
 and direction. For neutral pion momentum we assume an energy spread of 
$ 5 \%\over \sqrt{E [GeV]}$. For the $\theta$ and $\phi$ angular    spread
we assume  $ {1 \over 3}  {2 \pi \over 1800}$.  These neutral pion
resolutions can be    achieved with a 15\% energy error and a
2$\pi$/1800 direction error in the gammas   resulting from the $\pi^0$
decays. These resolutions have been approximately verified with a {\tt
SIMDET}~\cite{Pohl:2002vk}, a parametric Monte Carlo program for a
TESLA detector~\cite{Behnke:2001qq}. In Ref.\cite{Bower:2002zx} the 
method of reconstruction of 
replacement $\tau^{\pm}$ lepton rest frames was proposed. We will use 
this method here as well.
We boost the $\pi^+,\pi^0,\pi^-,\pi^0$ momenta to the respective 
replacement $\tau^{\pm}$ lepton rest frames.
The $\pi^{\pm}$ energies defined this way are used in  the 
$y_{1}$ and $y_{2}$ energy difference cuts.

When we  use the selection cuts $y_{1}$
and $y_{2}$ and  the replacement $\tau^{\pm}$ rest
frames as well as smearing    the $\pi^{\pm}$ momenta, we obtain    the
results shown in Fig.~\ref{acosm2}. We see that the effects to be
measured diminish but remain clearly visible.

\section*{Summary}

We have studied the possibility of distinguishing    a scalar from a
pseudoscalar couplings of light Higgs boson  to fermions   using its decay
to a pair of $\tau^{+}\tau^{-}$ leptons   and their subsequent decays to
$\tau^\pm \to \pi^\pm \bar{\nu}_{\tau}(\nu_{\tau})$ and
$\tau^\pm \to \rho^\pm \bar{\nu}_{\tau}(\nu_{\tau})$; 
$\rho^\pm \to \pi^\pm \pi^0$.

The beamstrahlung effect in
reconstruction of the  Higgs boson four-momentum degraded the method
of measuring the Higgs boson parity using the decay chain $H \to
\tau^+\tau^-$, $\tau^\pm \to \pi^\pm \bar{\nu}_{\tau}(\nu_{\tau})$ in
a decisive way. Therefore, there is little hope for  the  elegant method to
check Higgs boson parity using its decay to $\tau^\pm\to \pi^\pm
\bar{\nu}_{\tau}(\nu_{\tau})$,  whatever the luminosity of the future
Linear Collider is assumed.

For $\tau^\pm \to \rho^\pm \bar{\nu}_{\tau}(\nu_{\tau})$ decay,
we have discussed an observable which is very promising. Using
reasonable assumptions about the ${\cal SM}$ production cross section
and about the measurement resolutions we find that    with $500$
$fb^{-1}$ of luminosity at a $500$ $GeV$ $e^+e^-$ linear collider   the
${\cal CP}$ of a $120$ $GeV$ Higgs boson can be measured to a confidence level
greater    than $95\%$. We emphasize that the technique   is both model
independent and independent of the Higgs boson production    mechanism and
depends only on good measurements of the Higgs boson decay products.   Thus,
this method may be applicable to other production modes including
those available at proton colliders as well as at electron colliders.

\section*{Acknowledgments}
It is a pleasure to thank Gary Bower, Tomasz Pierzcha\l a
and  Zbigniew W\c as, with whom the work reported here was performed.
I wish to thank Marek Je\.zabek for giving me the opportunity to give
this talk at the Cracow Epiphany Conference on Heavy Flavors.  This
work is partly supported by the Polish State Committee for  Scientific
Research (KBN) grants Nos 5P03B09320, 2P03B00122.

\providecommand{\href}[2]{#2}

\begingroup\begin{thebibliography}{10}

\bibitem{Dell'Aquila:1988rx}
J.~R. Dell'Aquila and C.~A. Nelson, {\em Nucl. Phys.} {\bf B320} (1989) 61.

\bibitem{Dell'Aquila:fe}
J.~R. Dell'Aquila and C.~A. Nelson, {\em Nucl. Phys.} {\bf B320} (1989) 86.

\bibitem{Barger1994}
V.~Barger, K.~Cheung, A.~Djouadi, B.~A. Kniehl, and P.~M. Zerwas, {\em Phys.
  Rev.} {\bf D49} (1994) 79,
  \href{http://arXiv.org/abs/hep-ph/9306270}{{\tt hep-ph/9306270}}.

\bibitem{Hagiwara1994}
K.~Hagiwara and M.~Stong, {\em Z. Phys.} {\bf C62} (1994) 99,
  \href{http://arXiv.org/abs/hep-ph/9309248}{{\tt hep-ph/9309248}}.

\bibitem{Skjold:1994qn}
A.~Skjold and P.~Osland, {\em Phys. Lett.} {\bf B329} (1994) 305,
  \href{http://arXiv.org/abs/hep-ph/9402358}{{\tt hep-ph/9402358}}.

\bibitem{Boe:1998kp}
C.~A. Boe, O.~M. Ogreid, P.~Osland, and J.~Z. Zhang, {\em Eur. Phys. J.} {\bf
  C9} (1999) 413, \href{http://arXiv.org/abs/hep-ph/9811505}{{\tt
  hep-ph/9811505}}.

\bibitem{Hagiwara2000}
K.~Hagiwara, S.~Ishihara, J.~Kamoshita, and B.~Kniehl, {\em Eur. Phys. J.} {\bf
  C14} (2000) 457, \href{http://arXiv.org/abs/hep-ph/0002043}{{\tt
  hep-ph/0002043}}.

\bibitem{Kramer:1994jn}
M.~Kramer, J.~H. K\"u{hn}, M.~L. Stong, and P.~M. Zerwas, {\em Z. Phys.} {\bf
  C64} (1994) 21,
\href{http://arXiv.org/abs/hep-ph/9404280}{{\tt hep-ph/9404280}}.

\bibitem{Grzadkowski:1995rx}
B.~Grz\c{a}dkowski and J.~F. Gunion, {\em Phys. Lett.} {\bf B350} (1995) 218,
  \href{http://arXiv.org/abs/hep-ph/9501339}{{\tt hep-ph/9501339}}.

\bibitem{Aguilar-Saavedra:2001rg}
J.~A. Aguilar-Saavedra {\em et al.}, {\it "TESLA Technical Design Report Part
  III: Physics at an $e^+e^-$ Linear Collider"},
  \href{http://www.arXiv.org/abs/hep-ph/0106315}{{\tt hep-ph/0106315}}.

\bibitem{Mele:1992kh}
B.~Mele and G.~Altarelli, {\em Phys. Lett.} {\bf B299} (1993)
345.

\bibitem{Jadach:1990mz}
S.~Jadach, J.~H. K\"{uhn}, and Z.~W\c{a}s, {\em Comput. Phys. Commun.} {\bf 64}
  (1990)
275.

\bibitem{Jezabek:1991qp}
M.~Je\.zabek, Z.~W\c{a}s, S.~Jadach, and J.~H. K\"{uhn}, {\em Comput. Phys.
  Commun.} {\bf 70} (1992)
69.

\bibitem{Jadach:1993hs}
S.~Jadach, Z.~W\c{a}s, R.~Decker, and J.~H. K\"{uhn}, {\em Comput. Phys.
  Commun.} {\bf 76} (1993)
361.

\bibitem{Pythia}
{T. Sjostrand} {\em et al.}, {\em Comput. Phys. Commun.} {\bf 135} (2001) 238,
  \href{http://www.arXiv.org/abs/hep-ph/0010017}{{\tt hep-ph/0010017}}.

\bibitem{Was:2002gv}
Z.~W\c{a}s and M.~Worek, {\em Acta Phys.\ Polon.} {\bf B33} (2002) 1875,
  \href{http://www.arXiv.org/abs/hep-ph/0202007}{{\tt hep-ph/0202007}}.

\bibitem{Bower:2002zx}
G.~R. Bower, T.~Pierzcha\l{a}, Z.~W\c{a}s, and M.~Worek, {\em Phys.\ Lett.}
  {\bf B543} (2002) 227--234,
  \href{http://www.arXiv.org/abs/hep-ph/0204292}{{\tt hep-ph/0204292}}.

\bibitem{Pierzchala:2001gc}
T.~Pierzcha\l{a}, E.~Richter-W\c{a}s, Z.~W\c{a}s, and M.~Worek, {\em Acta Phys.
  Polon.} {\bf B32} (2001) 1277,
\href{http://arXiv.org/abs/hep-ph/0101311}{{\tt hep-ph/0101311}}.

\bibitem{Worek:2001hn}
M.~Worek, {\em Acta Phys. Polon.} {\bf B32} (2001) 3803,
\href{http://www.arXiv.org/abs/hep-ph/0110228}{{\tt hep-ph/0110228}}.

\bibitem{Kuhn:1982di}
J.~H. K\"{uhn} and F.~Wagner, {\em Nucl. Phys.} {\bf B236} (1984)
16.

\bibitem{Abe:2001gc}
{ACFA Linear Collider Working Group} Collaboration, {\it "Particle physics
  experiments at JLC"},
\href{http://arXiv.org/abs/hep-ph/0109166}{{\tt hep-ph/0109166}}.

\bibitem{Hagiwara:2002fs}
{Particle Data Group} Collaboration, K.~Hagiwara {\em et al.}, {\em Phys. Rev.}
  {\bf D66} (2002) 010001.

\bibitem{Pohl:2002vk}
M.~Pohl and H.~J. Schreiber, {\it " SIMDET - Version 4: A parametric Monte
  Carlo for a TESLA detector"},
  \href{http://www.arXiv.org/abs/hep-ex/0206009}{{\tt hep-ex/0206009}}.

\bibitem{Behnke:2001qq}
T.~Behnke, S.~Bertolucci, R.~D. Heuer, and R.~Settles, {\it "TESLA Technical
  Design Report Part IV: A detector for TESLA"},
  \href{http://www.arXiv.org/abs/DESY-01-011}{{\tt DESY-01-011}}.

\end{thebibliography}\endgroup
\end{document}